\documentstyle[sprocl]{article}

\def\fracerr#1{\ifmmode 
                    \frac{\delta #1}{#1}
               \else
                    \mbox{${\delta #1}/{#1}$}
               \fi}

\def\D0{D\O }

\def\Missing#1#2{{\rm {\mbox{$#1\kern-0.57em\raise0.19ex\hbox{/}_{#2}$}}\ }}

\def\vMissing#1#2{\rm{\ifmmode
            \vec{#1}\kern-0.57em\raise.19ex\hbox{/}_{#2}
         \else
            {{\mbox{$\vec{#1}\kern-0.57em\raise.19ex\hbox{/}_{#2}$}}\ }
         \fi}}

\def\MEt{\Missing{E}{T}}

\def\St{\ifhmode {$S_T$\ }\else{S_T}\fi}

\catcode`\@=11\relax
\newwrite\@unused
\def\typeout#1{{\let\protect\string\immediate\write\@unused{#1}}}
\def\figurepath{[]}

\def\@nnil{\@nil}
\def\@empty{}
\def\@psdonoop#1\@@#2#3{}
\def\@psdo#1:=#2\do#3{\edef\@psdotmp{#2}\ifx\@psdotmp\@empty \else
    \expandafter\@psdoloop#2,\@nil,\@nil\@@#1{#3}\fi}
\def\@psdoloop#1,#2,#3\@@#4#5{\def#4{#1}\ifx #4\@nnil \else
       #5\def#4{#2}\ifx #4\@nnil \else#5\@ipsdoloop #3\@@#4{#5}\fi\fi}
\def\@ipsdoloop#1,#2\@@#3#4{\def#3{#1}\ifx #3\@nnil 
       \let\@nextwhile=\@psdonoop \else
      #4\relax\let\@nextwhile=\@ipsdoloop\fi\@nextwhile#2\@@#3{#4}}
\def\@tpsdo#1:=#2\do#3{\xdef\@psdotmp{#2}\ifx\@psdotmp\@empty \else
    \@tpsdoloop#2\@nil\@nil\@@#1{#3}\fi}
\def\@tpsdoloop#1#2\@@#3#4{\def#3{#1}\ifx #3\@nnil 
       \let\@nextwhile=\@psdonoop \else
      #4\relax\let\@nextwhile=\@tpsdoloop\fi\@nextwhile#2\@@#3{#4}}
\def\psdraft{
        \def\@psdraft{0}
}
\def\psfull{
        \def\@psdraft{100}
}
\psfull
\newif\if@prologfile
\newif\if@postlogfile
\newif\if@noisy
\def\pssilent{
        \@noisyfalse
}
\def\psnoisy{
        \@noisytrue
}
\psnoisy
\newif\if@bbllx
\newif\if@bblly
\newif\if@bburx
\newif\if@bbury
\newif\if@height
\newif\if@width
\newif\if@rheight
\newif\if@rwidth
\newif\if@clip
\newif\if@verbose
\def\@p@@sclip#1{\@cliptrue}

\def\@p@@sfile#1{\def\@p@sfile{null}%
                \openin1=#1
                \ifeof1\closein1%
                       \openin1=\figurepath#1
                        \ifeof1\typeout{Error, File #1 not found}
                        \else\closein1
                            \edef\@p@sfile{\figurepath#1}%
                        \fi%
                 \else\closein1%
                       \def\@p@sfile{#1}%
                 \fi}
\def\@p@@sfigure#1{\def\@p@sfile{null}%
                \openin1=#1
                \ifeof1\closein1%
                       \openin1=\figurepath#1
                        \ifeof1\typeout{Error, File #1 not found}
                        \else\closein1
                            \def\@p@sfile{\figurepath#1}%
                        \fi%
                 \else\closein1%
                       \def\@p@sfile{#1}%
                 \fi}

\def\@p@@sbbllx#1{
                \@bbllxtrue
                \dimen100=#1
                \edef\@p@sbbllx{\number\dimen100}
}
\def\@p@@sbblly#1{
                \@bbllytrue
                \dimen100=#1
                \edef\@p@sbblly{\number\dimen100}
}
\def\@p@@sbburx#1{
                \@bburxtrue
                \dimen100=#1
                \edef\@p@sbburx{\number\dimen100}
}
\def\@p@@sbbury#1{
                \@bburytrue
                \dimen100=#1
                \edef\@p@sbbury{\number\dimen100}
}
\def\@p@@sheight#1{
                \@heighttrue
                \dimen100=#1
                \edef\@p@sheight{\number\dimen100}
}
\def\@p@@swidth#1{
                \@widthtrue
                \dimen100=#1
                \edef\@p@swidth{\number\dimen100}
}
\def\@p@@srheight#1{
                \@rheighttrue
                \dimen100=#1
                \edef\@p@srheight{\number\dimen100}
}
\def\@p@@srwidth#1{
                \@rwidthtrue
                \dimen100=#1
                \edef\@p@srwidth{\number\dimen100}
}
\def\@p@@ssilent#1{ 
                \@verbosefalse
}
\def\@p@@sprolog#1{\@prologfiletrue\def\@prologfileval{#1}}
\def\@p@@spostlog#1{\@postlogfiletrue\def\@postlogfileval{#1}}
\def\@cs@name#1{\csname #1\endcsname}
\def\@setparms#1=#2,{\@cs@name{@p@@s#1}{#2}}
\def\ps@init@parms{
                \@bbllxfalse \@bbllyfalse
                \@bburxfalse \@bburyfalse
                \@heightfalse \@widthfalse
                \@rheightfalse \@rwidthfalse
                \def\@p@sbbllx{}\def\@p@sbblly{}
                \def\@p@sbburx{}\def\@p@sbbury{}
                \def\@p@sheight{}\def\@p@swidth{}
                \def\@p@srheight{}\def\@p@srwidth{}
                \def\@p@sfile{}
                \def\@p@scost{10}
                \def\@sc{}
                \@prologfilefalse
                \@postlogfilefalse
                \@clipfalse
                \if@noisy
                        \@verbosetrue
                \else
                        \@verbosefalse
                \fi
}
\def\parse@ps@parms#1{
                \@psdo\@psfiga:=#1\do
                   {\expandafter\@setparms\@psfiga,}}
\newif\ifno@bb
\newif\ifnot@eof
\newread\ps@stream
\def\bb@missing{
        \if@verbose{
                \typeout{psfig: searching \@p@sfile \space  for bounding box}
        }\fi
        \openin\ps@stream=\@p@sfile
        \no@bbtrue
        \not@eoftrue
        \catcode`\%=12
        \loop
                \read\ps@stream to \line@in
                \global\toks200=\expandafter{\line@in}
                \ifeof\ps@stream \not@eoffalse \fi
                \@bbtest{\toks200}
                \if@bbmatch\not@eoffalse\expandafter\bb@cull\the\toks200\fi
        \ifnot@eof \repeat
        \catcode`\%=14
}       
\catcode`\%=12
\newif\if@bbmatch
\def\@bbtest#1{\expandafter\@a@\the#1
\long\def\@a@#1
\long\def\bb@cull#1 #2 #3 #4 #5 {
        \dimen100=#2 bp\edef\@p@sbbllx{\number\dimen100}
        \dimen100=#3 bp\edef\@p@sbblly{\number\dimen100}
        \dimen100=#4 bp\edef\@p@sbburx{\number\dimen100}
        \dimen100=#5 bp\edef\@p@sbbury{\number\dimen100}
        \no@bbfalse
}
\catcode`\%=14
\def\compute@bb{
                \no@bbfalse
                \if@bbllx \else \no@bbtrue \fi
                \if@bblly \else \no@bbtrue \fi
                \if@bburx \else \no@bbtrue \fi
                \if@bbury \else \no@bbtrue \fi
                \ifno@bb \bb@missing \fi
                \ifno@bb \typeout{FATAL ERROR: no bb supplied or found}
                        \no-bb-error
                \fi
                \count203=\@p@sbburx
                \count204=\@p@sbbury
                \advance\count203 by -\@p@sbbllx
                \advance\count204 by -\@p@sbblly
                \edef\@bbw{\number\count203}
                \edef\@bbh{\number\count204}
}
\def\in@hundreds#1#2#3{\count240=#2 \count241=#3
                     \count100=\count240        
                     \divide\count100 by \count241
                     \count101=\count100
                     \multiply\count101 by \count241
                     \advance\count240 by -\count101
                     \multiply\count240 by 10
                     \count101=\count240        
                     \divide\count101 by \count241
                     \count102=\count101
                     \multiply\count102 by \count241
                     \advance\count240 by -\count102
                     \multiply\count240 by 10
                     \count102=\count240        
                     \divide\count102 by \count241
                     \count200=#1\count205=0
                     \count201=\count200
                        \multiply\count201 by \count100
                        \advance\count205 by \count201
                     \count201=\count200
                        \divide\count201 by 10
                        \multiply\count201 by \count101
                        \advance\count205 by \count201
                     \count201=\count200
                        \divide\count201 by 100
                        \multiply\count201 by \count102
                        \advance\count205 by \count201
                     \edef\@result{\number\count205}
}
\def\compute@wfromh{
                \in@hundreds{\@p@sheight}{\@bbw}{\@bbh}
                \edef\@p@swidth{\@result}
}
\def\compute@hfromw{
                \in@hundreds{\@p@swidth}{\@bbh}{\@bbw}
                \edef\@p@sheight{\@result}
}
\def\compute@handw{
                \if@height 
                        \if@width
                        \else
                                \compute@wfromh
                        \fi
                \else 
                        \if@width
                                \compute@hfromw
                        \else
                                \edef\@p@sheight{\@bbh}
                                \edef\@p@swidth{\@bbw}
                        \fi
                \fi
}
\def\compute@resv{
                \if@rheight \else \edef\@p@srheight{\@p@sheight} \fi
                \if@rwidth \else \edef\@p@srwidth{\@p@swidth} \fi
}
\def\compute@sizes{
        \compute@bb
        \compute@handw
        \compute@resv
}
\def\psfig#1{\vbox {
        \ps@init@parms
        \parse@ps@parms{#1}
        \compute@sizes
        \ifnum\@p@scost<\@psdraft{
                \if@verbose{
                        \typeout{psfig: including \@p@sfile \space }
                }\fi
                \special{ps::[begin]    \@p@swidth \space \@p@sheight \space
                                \@p@sbbllx \space \@p@sbblly \space
                                \@p@sbburx \space \@p@sbbury \space
                                startTexFig \space }
                \if@clip{
                        \if@verbose{
                                \typeout{(clip)}
                        }\fi
                        \special{ps:: doclip \space }
                }\fi
                \if@prologfile
                    \special{ps: plotfile \@prologfileval \space } \fi
                \special{ps: plotfile \@p@sfile \space }
                \if@postlogfile
                    \special{ps: plotfile \@postlogfileval \space } \fi
                \special{ps::[end] endTexFig \space }
                \vbox to \@p@srheight true sp{
                        \hbox to \@p@srwidth true sp{
                                \hss
                        }
                \vss
                }
        }\else{
                \vbox to \@p@srheight true sp{
                \vss
                        \hbox to \@p@srwidth true sp{
                                \hss
                                \if@verbose{
                                        \@p@sfile
                                }\fi
                                \hss
                        }
                \vss
                }
        }\fi
}}
\def\psglobal{\typeout{psfig: PSGLOBAL is OBSOLETE; use psprint -m instead}}
\catcode`\@=12\relax

\bibliographystyle{unsrt} 

\arraycolsep1.5pt

\def\Journal#1#2#3#4{{#1} {\bf #2}, #3 (#4)}

\def\NCA{\em Nuovo Cimento}
\def\NIM{\em Nucl. Instrum. Methods}
\def\NIMA{{\em Nucl. Instrum. Methods} A}
\def\NPB{{\em Nucl. Phys.} B}
\def\PLB{{\em Phys. Lett.}  B}
\def\PRL{\em Phys. Rev. Lett.}
\def\PRD{{\em Phys. Rev.} D}
\def\ZPC{{\em Z. Phys.} C}

\def\epp{\epsilon^{\prime}}
\def\vep{\varepsilon}
\def\ppg{\pi^+\pi^-\gamma}
\def\ko{K^0}
\def\kb{\bar{K^0}}
\def\al{\alpha}
\def\ab{\bar{\alpha}}
\def\be{\begin{equation}}
\def\ee{\end{equation}}
\def\bea{\begin{eqnarray}}
\def\eea{\end{eqnarray}}
\def\CPbar{\hbox{{\rm CP}\hskip-1.80em{/}}}
\def\nub{\overline{\nu}}
\def\num{\nu_{\mu}}
\def\numb{\overline{\nu}_{\mu}}
\def\nue{\nu_{e}}
\def\nueb{\overline{\nu}_{e}}
\newcommand{\ubar}{\overline{u}}        
\newcommand{\dbar}{\overline{d}}
\newcommand{\qbar}{\overline{q}}
\newcommand{\alfs}{\mbox{$\alpha_s$}}
\newcommand{\asop}{\mbox{$\frac{\alpha_s}{\pi}$}}
\newcommand{\qsq}{\mbox{$Q^2$}}
\newcommand{\qnsq}{\mbox{$Q_0^2$}}
\newcommand{\mztwo}{\mbox{$M_Z^2$}}
\newcommand{\sinttw}{\mbox{$\sin^2\theta_W$}}
\newcommand{\stw}{\mbox{$\sin^2\theta_W$}}
\newcommand{\cotttw}{\mbox{$\cot^2\theta_W$}}
\newcommand{\sinftw}{\mbox{$\sin^4\theta_W$}}
\newcommand{\rnu}{\mbox{$R^{\nu}$}}
\newcommand{\rnub}{\mbox{$R^{{\overline \nu}}$}}
\newcommand{\lmsb}{\mbox{$\Lambda_{\overline{MS}}$}}
\newcommand{\st}{\scriptstyle}
\newcommand{\sst}{\scriptscriptstyle}
\newcommand{\mco}{\multicolumn}
\newcommand{\ra}{\rightarrow}
\newcommand{\vp}{{\bf p}}
\newcommand{\rmt}{\rm\textstyle}
\newcommand{\rms}{\rm\scriptstyle}
\newcommand{\nunuc}{\mbox{$\nu$N}}
\newcommand{\rmeas}[1]{\mbox{$R_{\rms meas}^{#1}$}}
\newcommand{\rmeasm}{R_{\rms meas}}
\newcommand{\mw}{\mbox{$M_W$}}
\newcommand{\mz}{\mbox{$M_Z$}}
\newcommand{\mwmw}{\mbox{$M_W^2$}}
\newcommand{\mzmz}{\mbox{$M_Z^2$}}
\newcommand{\mtop}{\mbox{$M_{\rms top}$}}
\newcommand{\mt}{\mbox{$M_{\rms top}$}}
\newcommand{\mhiggs}{\mbox{$M_{\rms Higgs}$}}

\begin{document}

\title{PRECISION MEASUREMENT OF $\stw$ FROM $\nu-N$ SCATTERING AT NuTeV
AND DIRECT MEASUREMENTS OF $M_{W}$}

\author{Jaehoon Yu\\
(for the NuTeV, D\O\, and CDF Collaborations)}

\address{MS309, FNAL, P.O.Box 500, Batavia,\\ IL 60510, USA\\E-mail: yu@fnal.gov} 


\maketitle\abstracts{ 
We present the preliminary result of $\stw$ from $\nu-N$ deep inelastic scattering
     experiment, NuTeV, at Fermilab.
This measurement of $\stw$ comes from measuring the Paschos-Wolfenstein
     parameter ${\cal R}^{-}=(\sigma_{NC}^{\nu}-\sigma_{NC}^{\nub})/(\sigma_{CC}^{\nu}-\sigma_{CC}^{\nub})$, 
     using separate beams of $\nu$ and $\nub$, utilizing the SSQT.
The resulting value of $sin^{2}\theta_{W}^{(on-shell)}$ is 
     $0.2253\pm0.0019({\rm stat})\pm0.0010({\rm syst})$.
This value is equivalent to the mass of the W boson, 
$M_{W}=80.26\pm0.11 {\rm GeV/c^{2}}$.
We also summarize the direct measurements of $M_{W}$ from the Tevatron 
     $\overline{p}p$ collider experiments, D\O\ and CDF.
Combining these two direct measurements yields 
$M_{W}=80.37\pm0.08 {\rm GeV/c^{2}}$.
}

\section{Introduction}
Mass of the $W$ boson ($M_{W}$) is a fundamental parameter in the electroweak 
       sector of the standard model (SM).
The parameters, Fermi constant ($G_{F}$), the fine structure constant
       ($\alpha$), mass of the $Z$ boson ($M_{Z}$), and electroweak 
       radiative correction $\delta r$ are expressed in terms of $M_{W}$.
Among these parameters $G_{F}$, $\alpha_{EM}$, and $M_{Z}$ are measured to
       very high precision.
Thus the measurement of $M_{W}$ can be used to constrain the mass of the
       standard model Higgs bosons, $M_{H}$, together with the measured
       top quark mass via radiative corrections.
In addition, since the radiative corrections takes modification with
       an introduction of new particles, precision measurement of $M_{W}$
       also provides constraint to new physics.

In this paper, we present the preliminary result of $\stw$ measurement 
       and the resulting mass of the $W$ boson from $\nu-N$ DIS experiment, 
       NuTeV, at Fermilab.
We also summarize the direct measurements of $M_{W}$ from the Tevatron 
       collider experiments, D\O\ and CDF. 
\section{Measurement of $\stw$ in $\nu-N$ DIS}\label{ss:nun-stw}

In tree level calculations, the electroweak mixing angle, $\stw$, appears
       in neutral current interactions of neutrinos and the calculations
       have process dependent radiative corrections.
The measurement of $\stw$ in $\nu-N$ scattering provides an indirect measure
       of $M_{W}$ via electroweak radiative corrections.
In addition, since the systematics are uncorrelated to the direct measurements
       of $M_{W}$ and the measurement can be done very precisely, the impact
       of the $\nu-N$ measurement is similar to those from the direct
       measurements.

Moreover, the comparisons of $M_{W}$ with those from direct measurements,
       $Z_{0}$ line shape, and $sin^{2}\theta_{W}^{\rm eff}$ provide 
       sensitivity to new physics.
Carrying out a model independent interpretation of $\stw$ measured from 
       $\nu-N$ DIS
       probes light quark couplings which is an extension of the standard
       model.
The measurement is also sensitive to new heavy particles, such as extra $Z$ 
       bosons which affects neutral current couplings.
In addition, comparisons of measured ratios of the neutral (NC) to 
       charged current (CC) cross sections
       to predictions without neutrino oscillations provide probes to
       neutrino oscillations.
\subsection{Previous Measurement of $\nu-N$ DIS}\label{subs:nun-prev}
Since the cross section of CC interactions of neutrinos is
   proportional to weak isospins ($I_{Weak}^{(3)}$) while that for 
   NC interactions is proportional to 
   ($I_{Weak}^{(3)}-Q_{EM}\stw$), the ratio of the CC to NC cross
   sections is proportional to $\stw$ as expressed in the Llewellyn-Smith 
   fomular~\cite{th:llsmith}:
\begin{eqnarray}\label{eq:rnu}
{\cal R}^{\nu({\overline{\nu}})}=\frac{\sigma_{NC}^{\nu({\overline{\nu}})}}
{\sigma_{CC}^{\nu({\overline{\nu}})}} = 
\rho^{2}\left( \frac{1}{2}-\sin^2\theta_W+
\frac{5}{9}\sin^4\theta_W
    \left( 1+\frac{\sigma^{\nub(\nu)}_{CC}}{\sigma^{\nu(\nub)}_{CC}}
            \right)\right).
\end{eqnarray}

The most important element in this measurement is separating NC from
     CC events.
This separation is obtained statistically by using the event length 
     variable.
The event length is defined for each event to be the number of counters 
     with energy deposition above 1/4 of that of a single muon.
The NC events have short length due to the absence of the muons in the event,
     while CC events are long.  
Thus, the experimentally measured ratio ${\cal R}_{meas}=N_{short}/N_{long}$, 
     represents the NC to CC cross section ratio, ${\cal R}^{\nu(\nub)}$.
We vary only the NC couplings in the detailed Monte Carlo (MC), until 
     ${\cal R}^{\nu(\nub)}$ from the MC matches with ${\cal R}_{meas}$.

The cross section model in the MC incorporates a leading order (LO) corrected 
     Quark-Parton-Model (QPM), using LO parton distribution functions
     from the CCFR structure function measurements.
The second part of the Monte Carlo simulates detailed detector and beam 
     effects, such as ``short'' CC events caused by low energy range-out 
     muons or by the muons exiting the detector fiducial volume,
     ``long'' NC events due to $\pi/K$ decays and punch through, 
     electron neutrino ($\nue$) CC events which appear short in the detector,
     and other detector effects that affect the event length variable.
The detailed MC also includes corrections;
     1) electromagnetic and electroweak radiative corrections,
     2) target isovector effects ($\sim 6\%$ for iron target),
     3) higher order QCD effects due to longitudinal structure functions, 
     $R_{L}$~\cite{ex:whitlow},
     and 4) effect due to heavy quark productions.

There are two major sources of systematic uncertainties in the CCFR
    measurements~\cite{ccfr:bjking,ex:ccfr-stw}.
The first is the theoretical uncertainty due to mass threshold effect in
    the CC production of charm-quark from the scattering off the sea quarks.
This effect is modeled by LO slow-rescaling.
The parameters in the slow rescaling model are measured by the CCFR 
    experiment~\cite{ccfr:dimu_charm}, 
    using events with two oppositely charged muons, where
    $m_{c}=1.31\pm0.24GeV/c^{2}$ 
    and $\kappa=0.37\pm0.05$.
The uncertainties in CC cross section calculations due to the 
      above two parameters result in $\delta$\stw$=0.0027$.

The second source is the lack of precise knowledge on $\nu_{e}$ flux in the 
    beam.
Approximately 80\% of the total $\nu_{e}$ in the CCFR neutrino beam, which is
       a mixture of $\nu$ and $\nub$, come from $K^{\pm}_{e3}$ decays.
The $\nu_{e}$ from this source is well constrained by observed 
       $K^{\pm}_{\mu2}$ spectra.
Remaining 16\% of $\nu_{e}$ come from neutral K decays, $K_{Le3}$, whose
    production cross section is known only to 20\%.
These two sources constitute 4.1\% uncertainty to $N_{\nu_{e}}$, based on 
    a Monte Carlo study.
A direct measurement, based on longitudinal shower development~\cite{ccfr:eta3},
    gives 3.5\% uncertainty.
Averaging these two uncertainties give 2.9\% uncertainty in $N_{\nu_{e}}$ which
       results in $\delta$\stw$=0.0015$.
Combining all the errors, the final CCFR result~\cite{ex:ccfr-stw} is
       $\stw=0.2236\pm0.0041$
which corresponds to on-shell mass of W boson
       $M_{W}=80.35\pm0.21 GeV/c^{2}$.

\subsection{Improvements in NuTeV}\label{subs:nutev}
Minimizing the two large systematic uncertainties requires 1) a technique
       insensitive to the sea quark distributions and 2) to minimize the number
       of electron neutrinos in the beam, especially resulting from $K_{L}$.
In order to minimize the uncertainty due to the charm quark production, the 
       NuTeV uses the Paschos-Wolfenstein parameter~\cite{th:paschos};
\begin{eqnarray}\label{eq:rmin}
{\cal R}^{-}=\frac{\sigma_{NC}^{\nu}-\sigma_{NC}^{\nub}}
{\sigma_{CC}^{\nu}-\sigma_{CC}^{\nub}}=\frac{R^{\nu}-rR^{\nub}}{1-r}
=(g_{L}^{2}-g_{R}^{2})=\rho\left(\frac{1}{2}-\stw \right)
\end{eqnarray}
Since $\sigma^{\nu q}=\sigma^{\nub \qbar}$ and 
 $\sigma^{\nub q}=\sigma^{\nu \qbar}$, the effect of scattering off the 
      sea 
      quarks cancels by taking the differences in the neutrino and antineutrino
      cross sections.
However, the measurement of this quantity is complicated due to the fact
      that the NC final states look identical for $\nu$ and $\nub$.
Thus, in order to use this relationship, one needs to be able to distinguish
    neutrino NC interactions from antineutrino interactions which can only be
    achieved by having prior knowledge on the beam. 
To achieve this discrimination, the NuTeV modified the beamline to
    use a Sign-Selected-Quadrupole-Train (SSQT) to select either 
    $\nu$ or $\nub$ beam at a given running period.

The uncertainty caused by the $\nu_{e}$ flux is minimized by
    adding a 7.8 mrad upward angle to the incident proton beam relative to
    the horizontal axis which directs to the detector.
This upward incident angle together with a dipole stationed immediately behind
    the production target causes the neutral
    secondaries (especially $K_{L}$), oppositely charge secondary mesons, 
    and remnant protons 
    to be directly absorbed into the dumps in the SSQT.
The remaining major source of $\nue/\nueb$ is $K^\pm_{e3}$ decays.
The fractional uncertainty on predictions of $\nue/\nueb$ from $K^\pm_{e3}$ 
    is $\approx1.5\%$ and is dominated by the uncertainty in $K^\pm_{e3}$
    branching ratio.

\subsection{Result of $\stw$ from NuTeV}
The extraction of $\stw$ in NuTeV is based on the data sample of 1.3 million 
    neutrino and 0.3 million antineutrino events passing the following 
    set of cuts;
    hadronic energy measured in the calorimeter above 20GeV to ensure 
    full efficiency of triggers and vertex identification, and
    the location of the neutrino interaction must be 1) within the central
    2/3 of the calorimeter transverse dimension to ensure full acceptance,
    2) at least 0.4~m from the upstream end of the 
    calorimeter to minimize non-neutrino backgrounds, and
    3) at least 2.4~m from the downstream end to ensure sufficient calorimeter
    coverage for event length measurement.

\begin{figure}[tbp]
\begin{center}
\centerline{\psfig{file=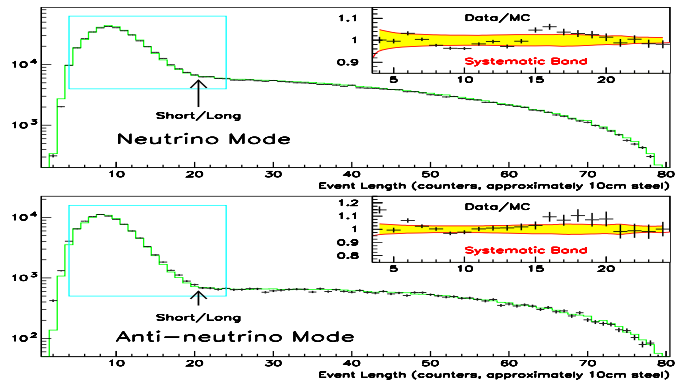,width=3.5in,height=2.0in}}
\caption[]{Comparisons of ``event length'' variable between data and MC for
neutrino (top) and antineutrino (bottom) modes.  The shaded areas in the
inset represent systematic uncertainties and the arrows indicate the cut value
for NC and CC distinctions.}
\label{fg:length}
\end{center}
\end{figure}

The ratios of NC candidates (short length) to CC candidates (long length), 
    ${\cal R}_{meas}^{\nu(\nub)}$, are
    $0.4198\pm0.008$ and $0.4215\pm0.0017$ for neutrino and antineutrino,
    respectively.
As it has been discussed in section~\ref{ss:nun-stw}, ${\cal R}_{meas}$ is
    then related to detailed MC predictions.
The detector MC in NuTeV has various improvements to reduce previous 
      experimental 
      systematic uncertainties further, utilizing muons from the upstream
      neutrino interactions and the information obtained from extensive
      in-situ and continuous calibration beam.
Figure~\ref{fg:length} shows the comparisons of the event length variables
    between data and MC for neutrino (top) and antineutrino modes(bottom),
    demonstrating good agreements in length distributions, especially in
    the cut value region.

To extract $\stw$ in NuTeV, we form a linear combination of 
     ${\cal R}^{\nu}_{meas}$ and ${\cal R}^{\nub}_{meas}$;
\begin{eqnarray}
{\cal R}^{-}_{meas}\equiv {\cal R}^{\nu}_{meas}-\alpha {\cal R}^{\nub}_{meas},
\end{eqnarray}
where $\alpha$ is determined using the MC such that ${\cal R}^{-}_{meas}$ is
     insensitive to small changes in the CC cross sections due to charm mass
     threshold effect.
For the measurement, the value of $\alpha$ is found to be 0.5136.
This technique is essentially employing the expression in the third 
     term in Eq.~\ref{eq:rmin}
     instead of the second term which requires separate background estimate
     and flux normalizations.
This technique cancels out large number of systematics by taking the ratios
     separately in neutrino and antineutrino modes, at the same time largely
     canceling the the uncertainties related to charm quark production from
     the scattering off the sea.
The remaining small undertainty due to heavy quark production comes from the
      scattering off the d-valence quark which is Cabbibo suppressed.

\begin{table}[tbp]
\caption[]{Summary of uncertainties in NuTeV $\stw$ measurements.}
\label{tb:syst}
\begin{center}
\begin{tabular}{|r|c|}
\hline
  SOURCE OF UNCERTAINTY & $\delta\sin^2\theta_W$ \\
\hline
  { {\em Statistics}: \hfill {Data} } & {\bf{0.00188}}  \\
  { Monte Carlo } & {0.00028} \\ \hline
  { \bf TOTAL STATISTICS \hfill } & {0.00190} \\ \hline\hline
  { $\nu_e/\nub_e$  } & {0.00045} \\
  { Energy Measurement} & {0.00051} \\
  { Event Length} & {0.00036} \\
  \hline
  { \bf TOTAL EXP. SYST. \hfill } & {0.00078} \\ \hline\hline
  { Radiative Corrections } & {0.00051} \\
  { Strange/Charm Sea } & {0.00036}  \\
  { Charm Mass } & {0.00009}  \\
  { $u/d$, $\ubar/\dbar$ } & {0.00027}  \\
  { Longitudinal Structure Function } & {0.00004}  \\ 
  { Higher Twist } & {0.00011} \\ \hline
  { \bf TOTAL PHYSICS MODEL \hfill } & {0.00070} \\  \hline\hline  
  { \bf TOTAL UNCERTAINTY \hfill } & {0.0022} \\  \hline
\end{tabular}
\end{center}
\end{table}
Table~\ref{tb:syst} summarizes various uncertainties in this measurement.
The single dominant uncertainty is the data statistical uncertainty whose
     magnitude is about twice as large as the total systematic uncertainty.
\subsection{ $M_{W}^{on-shell}$ from $\stw$}
The preliminary result from the NuTeV $\stw$ measurement in on-shell
renormalization scheme is :
\begin{eqnarray}\label{eq:stw}
sin^{2}\theta_{W}^{(on-shell)}=0.2253\pm0.0019({\rm stat})\pm0.0010({\rm syst})\\
-0.00142\times\left(\frac{M_{top}^{2}-(175GeV)^{2}}{(100GeV)^{2}} \right) \nonumber\\
+0.00048\times log_{e}\left( \frac{M_{H}}{150GeV} \right) \nonumber.
\end{eqnarray}
The small residual dependence of this result on $M_{top}$ and $M_{H}$ comes
    from the leading terms in the electroweak radiative 
    corrections~\cite{th:rad}.
Since within the on-shell renormalization scheme, the $\stw$ is related 
    to $M_{W}$
    and $M_{Z}$ together with the standard model prediction of $\rho$,
$\stw\equiv 1-\frac{M_{W}^{2}}{M_{Z}^{2}}$, this result is equivalent to;
\begin{eqnarray}
M_{W}=80.26\pm0.10({\rm stat})\pm0.05({\rm syst})\\
+0.073\times \left( \frac{M_{top}^{2}-(175GeV)^{2}}{(100GeV)^{2}} \right) \nonumber\\
-0.025\times log_{e}\left( \frac{M_{H}}{150GeV} \right) \nonumber.
\end{eqnarray}
\begin{center}
\begin{figure}[tbp]
\centerline{\psfig{figure=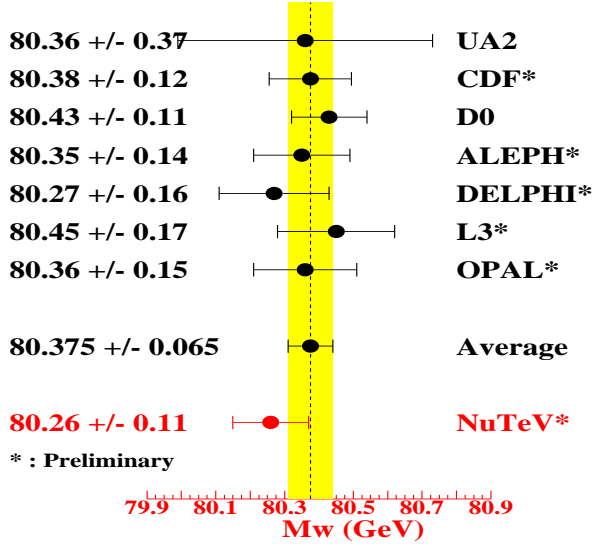,width=4.5in,height=3.5in}}
\vskip -0.2in
\caption[]{Various measurements of $M_{W}$.}
\label{fg:world-mw}
\end{figure}
\end{center}
\begin{table}[tpb]
\caption[]{Summary of data sample and analyses cuts of the direct $M_{W}$ 
measurements.}
\label{tb:col-cuts}
\begin{center}
\begin{tabular}{|c|c|c|}\hline\hline
Parameters & CDF & D\O\ \\ \hline\hline
Decay Channel & $W\rightarrow \mu+\nu$ &  $W\rightarrow e+\nu$\\ \hline
$\int {\cal L}dt $ & $90pb^{-1}$ & $82pb^{-1}$\\ \hline
Number of W & 21,000 & 28,000 \\ \hline
Number of Z & 1,400 & 2,200 \\ \hline
 $P^{l}_{T}$ cut & $25<P_{T}^{\mu}<60$GeV & $E_{T}^{e}>25$GeV \\ \hline
 \MEt Cut & $25<\MEt<60$ GeV & $25<\MEt$ \\ \hline
 Recoil $U_{T}$ cut & $U_{T}<20$ & $U_{T}<15$ \\ \hline
Lepton Rapidity cut & $\vert\eta_{\mu}\vert\leq 1 $ 
                    & $\vert\eta_{e}\vert\leq 1 $ \\ \hline
Lepton Quality cuts & $\mu$ quality & e-quality \\
                    & MiP in Calorimeter & Isolation \\
                    & Cosmic removal & Shower shape\\
                    & Track Match & Track match\\ \hline
$M_{T}^{W}$ cut  & $50<M_{T}^{W}<110$GeV & $50<M_{T}^{W}<110$GeV \\ \hline\hline
\end{tabular}
\end{center}
\end{table}
Figure~\ref{fg:world-mw} shows world measurements of the mass of the W boson.
This result is about one standard deviation lower than the world average and
      is as precise as the measurements from the Tevatron collider
      experiments.
Thus this measurement of $\stw$ from the $\nu-N$ 
      scattering~\cite{ex:ksm-moriond} contributes 
      significantly
      to the determination of $M_{W}$ and thereby constraining the mass of
      standard model Higgs boson, $M_{H}$.
\section{Direct Measurements of $M_{W}$}
Since the effective center of mass energy of Tevatron $\overline{p}p$
    collider is higher than the mass of the W boson, $M_{W}$, real
    $W$'s are produced in the collider experiments at $\sqrt{s}=1.8$TeV.
The D\O\ and CDF experiments at Fermilab use leptonic decay modes of
    the $W$ bosons, $W\rightarrow e+\nu$ and $W\rightarrow \mu+\nu$
    for D\O\ and CDF, respectively.
The results presented in this paper are based on Tevatron collider run 
     Ib data combined with the Run Ia from both experiments.

The technique used in the direct measurements of $M_{W}$ in collider experiment
     relies on both the transverse momentum ($E_{T}^{l}$) spectra of 
     charged leptons
     resulting from the leptonic decays of the $W$'s or the transverse mass
     ($M_{T}^{W}$)  distributions computed using
     $E_{T}^{l}$ and missing momentem ($\MEt$) which infers the neutrino 
     momentum: $M_{T}^{W}=\sqrt{2\MEt E^{l}_{T}(1-cos\phi_{l\nu})}$.
Table\ref{tb:col-cuts} lists the data sample and the cuts used for the
     analyses.

The measured $E_{T}^{l}$ and $M_{T}^{W}$ distributions of the $W$ 
          candidates are then compared to the detailed MC simulations
          which incorporates detector effects and theoretical models.
Since $E_{T}^{l}$ and $M_{T}^{W}$ are sensitive to the detector 
          energy scale and resolutions, and the underlying events affects
          the recoil system, both collaborations use
          $Z\rightarrow ll$ events to measure these systematic effects.
The theoretical model incorporated in the detailed MC simulation includes
          various available next-to-leading order (NLO) parton distribution
          functions.
The momentum distributions of the $W$'s affect the $E_{T}^{l}$ and $M_{T}^{W}$
          and is modeled by a theoretical predictions~\cite{th:yuan} 
          with higher-twist
          effects to account for non-perturbative effects.
\begin{center}
\begin{figure}[tbp]
\vskip -2.in
\begin{minipage}{27.0cm}
\begin{picture}(30.0,150.0)(0.0,35.0)
\psfig{figure=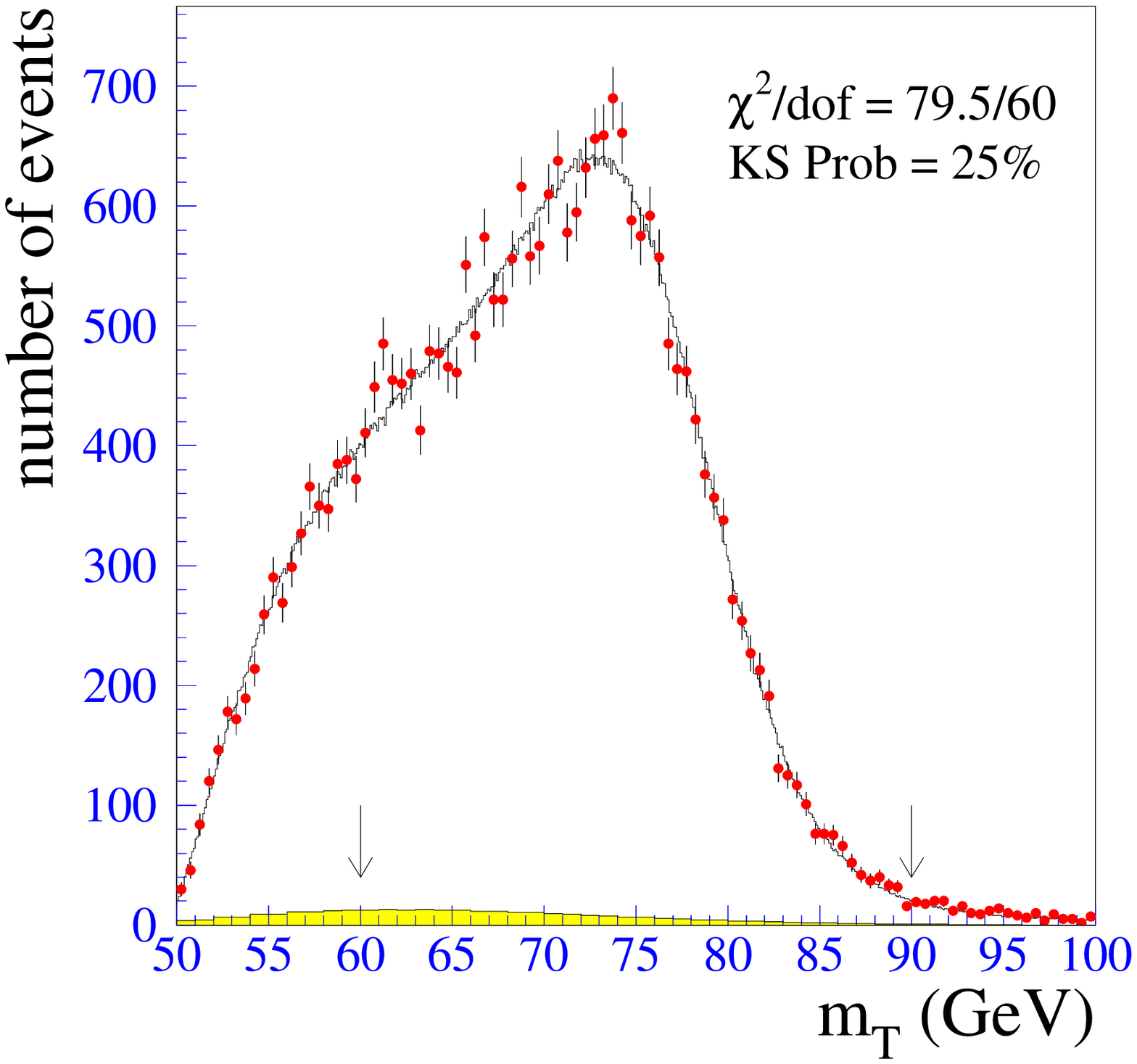,width=2.4in,height=2.6in}
\end{picture}
\begin{picture}(30.0,280.0)(-130.0,235.0)
\psfig{figure=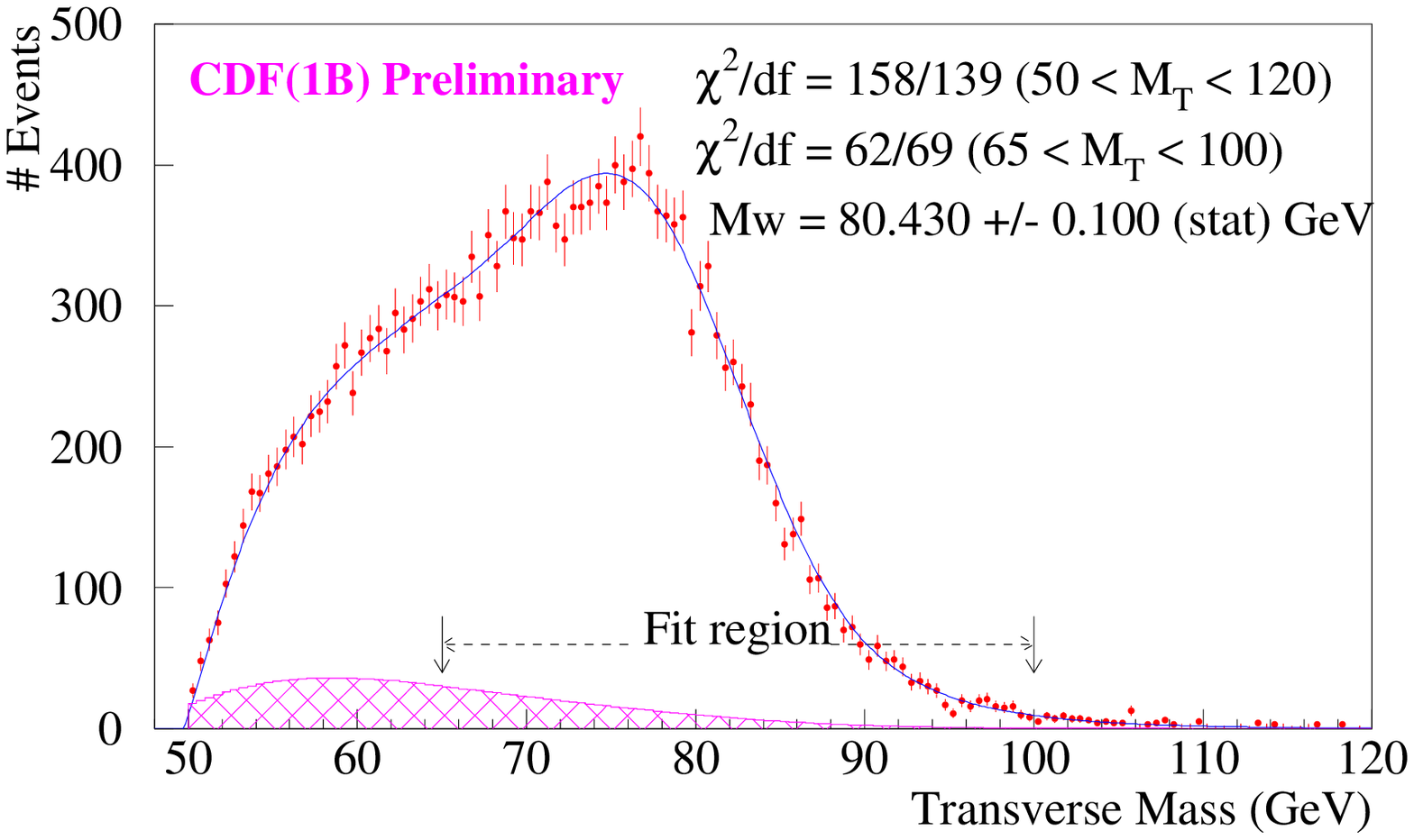,width=2.4in,height=4.075in}
\end{picture}
  \vspace{1.cm}
 \end{minipage}
\vspace{5pt}
\vskip -0.2in
\caption[]{Transverse mass distributions of $W\rightarrow e\nu(\mu\nu)$ 
events from D\O\ and CDF.}
\label{fg:mt-d0-cdf}
\end{figure}
\end{center}

Figure~\ref{fg:mt-d0-cdf} shows $M_{T}^{W}$ distributions from the D\O\
     and CDF experiments using the electron and muon decay channels, 
     respectively.
The lines in the plots are the best MC fit of $M_{W}$ to the data $M_{T}^{W}$
     distributions.
The same type of MC fits to $E_{T}^{l}$ distribution are also performed.
Combining the $M_{W}$ from both the fits in each experiment result in ;
\begin{eqnarray}\label{eq:d0-cdf-mw}
M_{W}=80.38\pm0.11 {\rm GeV/c^{2}}~~~(D\O)\\
M_{W}=80.43\pm0.12 ({\rm syst}) {\rm GeV/c^{2}}~~~(CDF).
\end{eqnarray}
Taking the weighted average of these two measurements together with
      UA2~\cite{ex:ua2-mw} results in ;
\begin{eqnarray}\label{eq:direct-mw}
M_{W}=80.37\pm0.08 {\rm GeV/c^{2}}.
\end{eqnarray}
Table~\ref{tb:col-syst} summarizes various uncertainties in these measurements
        from Run Ib data only.
\begin{table}[tpb]
\caption[]{Summary of uncertainties in direct $M_{W}$ measurements from the
D\O\ and CDF experiments~\cite{ex:tev-ib}.}\label{tb:col-syst}
\begin{center}
\begin{tabular}{|c|c|c|}\hline\hline
Source & $\delta M_{W}^{CDF}$ (MeV) & $\delta M_{W}^{D\O}$ (MeV) \\ \hline\hline
Statistical & 100 & 70 \\ \hline
Momentum \& Energy scale & 40 & 65 \\
Calorimeter linearity & - & 20 \\
Lepton $P_{T}$ resolution & 25 & 20 \\
Recoil Modeling & 90 & 40 \\
Input $P_{T}^{W}$ and PDF's & 50 & 25 \\
Radiative decays & 20 & 20 \\
Higher Order Corrections & 20 & - \\
Backgrounds & 25 & 10 \\
Lepton Angle Calibration & - & 30 \\
Momentum Fitting & 10 & - \\
Other errors & 20 & 20 \\ \hline
Total Systematic Uncertainty & 115 & 70 \\ \hline\hline
Total Uncertainty & 155 & 120 \\ \hline\hline
\end{tabular}
\end{center}
\end{table}
\section{Constraining Standard Model Higgs Mass}
\begin{figure}
\centerline{\psfig{figure=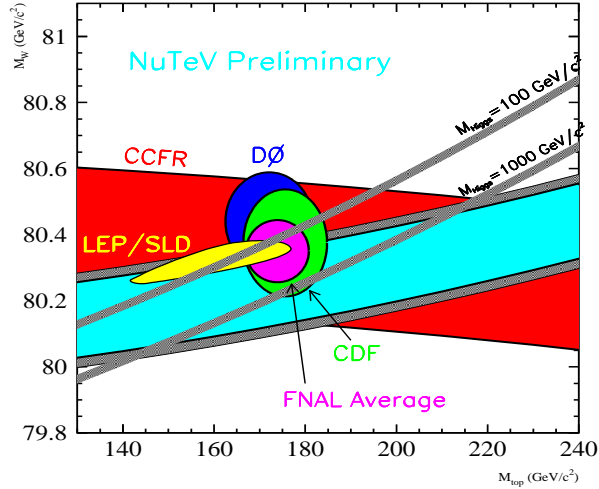,width=3.5in,height=3.0in}}
\caption[]{The 68\% confidence level contour from various measurements of 
$M_{W}$.}\label{fg:mt-mw}
\end{figure}
Using the measurements of $M_{W}$ and the relationship of $M_{W}$ with
    $M_{top}$ and $M_{H}$ via electroweak radiative corrections, one can
    constrain the standard model higgs mass.
Figure~\ref{fg:mt-mw} shows the 68\% confidence level contours from various 
    measurements.
Since the NuTeV does not measure $M_{top}$, the result from the NuTeV appears
    as a band in this plot.
Combining all hadron collider direct measurements with NuTeV preliminary 
      result yields the $M_{W}$~\cite{ex:cdf-ew-web} of;
\begin{eqnarray}
M_{W}=80.345\pm 0.055 {\rm GeV/c^{2}}.
\end{eqnarray}
This mean value of the $M_{W}$ and the current best measurement of $M_{top}$
      would constrain the standard model Higg mass to less than $\sim 500$GeV.
%
%
\section{Conclusions}
The NuTeV has finished taking its data in September 1997, accumulating 
      $3\times 10^{18}$ protons on target.
The NuTeV presented a preliminary result of $\stw$ measurement which is 
      equivalent to
      the mass of the W boson; $M_{W}=80.26\pm0.11~{\rm GeV/c^{2}}$.
This result is as precise as the direct measurements from the Tevatron collider
      experiments.
Further improvement in this analysis from the NuTeV experiment is
      modest because
      the result is already dominated by data statistical uncertainty.
Model independent interpretation of light quark couplings will come
      soon and the results from other analyses are also expected in the
      near future.

The two collider experiments are continuing their analyses of Run I data, to
      further improve the measurements of $M_{W}$.
The CDF experiment expects the total uncertainty reduce to $\sim90$ MeV, 
      by combining
      electron decay channel from Run Ib together with the muon decay data.
The D\O\ experiment expects the uncertainty of $\sim100$ MeV, by including the
      forward ($\vert\eta^{e}\vert>1$) electron data.

The direct measurements of $M_{W}$ are expected to improve dramatically in 
      Tevatron Run II with higher luminosity 
      and upgraded detectors.
The expected uncertainty in $M_{W}$ measurement is on the order of 40-50MeV 
      with the expected $\int {\cal L}dt=2fb^{-1}$ for each detector.
This improvement is expected not only due to improved statistics but also 
      due to various improvements in experimental and theoretical systematic 
      uncertainties, such as better measured detector resolution, recoil 
      $P_{T}$ due to increased $Z\rightarrow ll$ statistics, better measured
      $\Gamma_{W}$, better determined parton
      distribution functions, improved radiative decays, etc.

Benefitting from all these improvement, the Tevatron measurements of $M_{W}$
      together with the much more precisely measured $M_{top}$ would 
      further constrain the standard model Higgs mass, $M_{H}$.
Combining these indirect constraints on $M_{H}$ with direct searches 
      would enable the experiments either
      to find the Higgs bosons or to further constrain $M_{H}$ to narrow 
      down the regions to be searched in future machines, such as LHC.
%
%

\end{document}